\documentclass[reprint,showpacs,preprintnumbers,pre,superscriptaddress]{revtex4-1}
\usepackage{graphicx}
\begin{document}

\title{General neck condition for the limit shape of budding vesicles}
\author{Pan Yang}
\affiliation{Department of Physics, Beijing Normal University, Beijing 100875, China}
\affiliation{Applied Physics and Applied Mathematics Department, Columbia University, New York 10027, USA}
\author{Qiang Du}
\affiliation{Applied Physics and Applied Mathematics Department, Columbia University, New York 10027, USA}
\author{Z. C. Tu}\email[Corresponding author. Email: ]{tuzc@bnu.edu.cn}
\affiliation{Department of Physics, Beijing Normal University, Beijing 100875, China}

%\date{\today}

\begin{abstract}
The shape equation and linking conditions for a vesicle with two-phase domains are derived. We refine the conjecture on the general neck condition for the limit shape of a budding vesicle proposed by J\"{u}licher and Lipowsky [Phys. Rev. Lett. \textbf{70}, 2964 (1993); Phys. Rev. E \textbf{53}, 2670 (1996)], and then we use the shape equation and linking conditions to prove that this conjecture holds not only for axisymmetric budding vesicles, but also for asymmetric ones.
Our study reveals that the mean curvature at any point on the membrane segments adjacent to the neck satisfies the general neck condition for the limit shape of a budding vesicle when the length scale of the membrane segments is much larger than the characteristic size of the neck but still much smaller than the characteristic size of the vesicle.
\end{abstract}
\maketitle

\section{introduction}
In 1855 Virchow proposed his famous cell theory ``Omnis cellula e cellula", which means all cells come from pre-existing cells by division. As a crucial stage of cell division, cytokinesis is an orchestrated process that marks the beginning of a new cellular generation.
All cells are enclosed by plasma membranes which maintain the physical integrity of cells and regulate the intercellular exchange of matter and information. During cytokinesis, a contractile ring grows beneath the plasma membrane, which is a structure mainly composed of actin filaments and motor proteins. When constricted, the contractile ring generates a
force on the plasma membrane, and then partitions the cell into two daughters. Although the tension generated by the contractile ring has a great effect on cytokinesis, this force has still not been accurately measured because the spatial organization and motions of the components within the contractile ring are poorly characterized~\cite{Stachowiak14}.

Budding lipid vesicles have long been used as ideal models to mimic cytokinesis. The budding configurations are determined by several physical factors~\cite{Gozdz01,Harden05,GarciaSaez07,Seifert91,LipowskyPRL93,LipowskyPRE96,Roux05,Parthasarathy06,Dobereiner93,Gozdz01,Cox15,Shlomovitz08,Dorn16,Liu-Oster06} including the spontaneous curvature and bending elasticity of lipid membranes, as well as the line tension that reflects the constricting force of contractile ring. Therefore, the study of budding vesicles may provide a potential approach to measure the force induced by a contractile ring.
Seifert \emph{et al.}~\cite{Seifert91} investigated a budding vesicle in uniform phase without taking into account of the line tension.
By numerically optimizing Helfrich's free energy~\cite{Helfrich73} in the axisymmetric situation, Seifert and his co-workers found that the following neck condition
\begin{equation}\label{eq-neckcd4}
\frac{1}{R^\mathrm{I}}+\frac{1}{R^\mathrm{II}}=c_0
\end{equation}
holds for a limit shape consisting of two spheres connected by an infinitesimal neck, where $c_0$ is the spontaneous curvature of the lipid bilayer that constitutes the lipid vesicle. $R^\mathrm{I}$ and $R^\mathrm{II}$ represents the radius of two spheres, respectively. By doing a variation of Helfrich's free energy~\cite{Helfrich73} with an axisymmetric trial configuration consisting of two hemispheres connected by a catenoid-like surface, Fourcade \emph{et al.}~\cite{Fourcade94} analytically confirmed the numerical result obtained by Seifert and his co-workers. J\"{u}licher and Lipowsky subsequently found a more general neck condition for a budding vesicle with two-phase domains through numerical simulations where the bending energy of both domains, as well as the line tension of the separation boundary of two domains were involved~\cite{LipowskyPRL93,LipowskyPRE96}. They obtained the following neck condition
\begin{equation}
\frac{k_c^\mathrm{I}}{R^\mathrm{I}}+\frac{k_c^\mathrm{II}}{R^\mathrm{II}}=\frac{1}{2}(k_c^\mathrm{I}c_0^\mathrm{I}+k_c^\mathrm{II}c_0^\mathrm{II}+\gamma)
\end{equation}
for a limit shape consisting of two spheres connected by an infinitesimal neck. $R^\mathrm{I}$ and $R^\mathrm{II}$ represent the radius of two spheres, respectively. $k_c^\mathrm{I}$ and $k_c^\mathrm{II}$ are bending moduli of two domains, respectively. $c_0^\mathrm{I}$ and $c_0^\mathrm{II}$ are the spontaneous curvatures of the two domains, respectively. $\gamma$ represents the line tension of the separation curve between two domains. They also analytically confirmed this neck condition~\cite{LipowskyPRE96} using the method
developed by Fourcade and his co-workers. This relation degenerates to Eq. (\ref{eq-neckcd4}) for a budding vesicle with uniform phase where $k_c^\mathrm{I}=k_c^\mathrm{II}$, $c_0^\mathrm{I}=c_0^\mathrm{II}$ and $\gamma=0$.
Their numerical study further reveals that the limit shape consisting of two axisymmetric but nonspherical vesicles connected by an infinitesimal neck satisfy a general neck condition
\begin{equation}\label{eq-neckcd}
k_c^\mathrm{I}\left(2H^\mathrm{I}_{\epsilon}+c_0^\mathrm{I}\right)+k_c^\mathrm{II}\left(2H^\mathrm{II}_{\epsilon}+c_0^\mathrm{II}\right)+\gamma=0.
\end{equation}
%in the absence of volume constraint.
Here, $H_\epsilon^\mathrm{I}$ and $H_\epsilon^\mathrm{II}$ denote the mean curvatures at points in the two domains adjacent to the neck. Note that the sign of mean curvature here is opposite to those defined in the work by J\"{u}licher and Lipowsky~\cite{LipowskyPRE96}. In order to indicate the degree of adjacency
measured by a parameter $\epsilon$, we add $\epsilon$ as the subscript of the mean curvature $H$.

The general condition (\ref{eq-neckcd}) is an elegant identity, which connects the spontaneous curvature, local mean curvature and line tension.
Although J\"{u}licher and Lipowsky merely verified this identity by using the special trial configuration consisting of two hemispheres connected by a catenoid-like surface, they conjectured that the neck condition for limit shapes is quite general, which at least holds for axisymmetric budding vesicles.
In this paper and to be consistent with the convention adopted in the literature   \cite{LipowskyPRL93,LipowskyPRE96},
the limit shape is generally defined as a configuration consisting of two individual sub-vesicles connected by an infinitesimal neck and the two domains appear to be tangentially ``kissing'' at a single point from the macroscopic view.
The conjecture has aroused a great deal of  studies~\cite{GompperPRE99,BaumgartN03,BaumgartBJ05,BaumgartEPL09,TuJPA04,TuJCTN08,DuJMB08,LipowskyACIS14,TuACIS14,SvetinaPRE14} on shape transitions of vesicles with two-phase domains. If such a conjecture is true, researchers may utilize the general neck condition as a remedy to overcome the aforementioned difficulty in the measurement of the force generated by the contractile ring during cytokinesis. However, it is still an open question whether the conjecture on general neck condition for limit shapes is true or false, even in the axisymmetric situation. In addition, J\"{u}licher and Lipowsky did not specify  the applicable  range of the neck condition. In other words, the meaning of ``adjacent to the neck"~\cite{LipowskyPRE96} remains unclear. In this paper we will theoretically prove that the ``general" neck condition (\ref{eq-neckcd}) is indeed quite universal, which is applicable not only for axisymmetric budding vesicles, but also for asymmetric ones. Furthermore, we offer a quantitative definition of the ``adjacency to the neck", i.e., a specification of
the characteristic length scale  $\epsilon$ that has appeared in Eq.~(\ref{eq-neckcd}).

The limit shape of a budding vesicle involves multiple spatial scales.
The first scale $l_v$ is the characteristic length of the vesicle.
For a vesicle with $l_v$ being the characteristic length, after substituting the corresponding mean curvature and Gaussian curvature into the shape equation of lipid vesicles obtained in Ref.~\cite{oyPRL87,oyPRA89}, one may find that the order of $l_v$ is in accordance with the smaller one among the reciprocal of spontaneous curvature and the ratio of surface tension to osmotic pressure. Thus for a vesicle with two-phase domains we may take
\begin{equation}\label{eq-lv}
l_v\simeq\min\{1/c_0^\mathrm{I},1/c_0^\mathrm{II},\lambda^\mathrm{I}/p,\lambda^\mathrm{II}/p\},
\end{equation}
where $\lambda^\alpha$ and $p$ are the surface tension of domain $\alpha~(=\mathrm{I},~\mathrm{II})$ and the osmotic pressure of the budding vesicle.
The second scale $l_n$ is the characteristic length of the infinitesimal neck, which is much smaller than $l_v$. Although the curvatures of the neck curve at different points on the neck take different values, we expect that all of them are of the same order of magnitude. For simplicity, we take
\begin{equation}\label{eq-ln}
l_n\simeq\frac{1}{\kappa_m}
\end{equation}
with $\kappa_m$ being the maximum curvature of the neck curve. Since $l_n\ll l_v$ in the limit shape, these two length parameters may be regarded as macroscopic and microscopic scales, respectively. There also exists an intermediate scale $l_i$ which may be constructed from the macroscopic and microscopic scales
\begin{equation}\label{eq-li}
l_i=\sqrt{l_nl_v}.
\end{equation}
Obviously this construction guarantees $l_n\ll l_i\ll l_v$ if $l_n\ll l_v$.

\begin{figure}[htp!]
\centerline{\includegraphics[width=7cm]{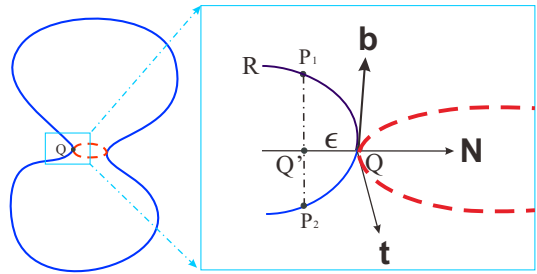}}\caption{\label{fig0-epsilon}Geometric definition of $\epsilon$ in Eq.~(\ref{eq-neckcd}).}
\end{figure}

Before depicting the length scale of $\epsilon$ in Eq.~(\ref{eq-neckcd}) which indicates the degree of proximity
 that a point approaches the neck, we first give a geometric definition of $\epsilon$.
As shown in Fig.~\ref{fig0-epsilon}, at any point $\mathrm{Q}$ on the neck (dashed line in the figure), the tangent vector, the normal vector and the binormal vector are denoted as ${\mathbf{t}}$, ${\mathbf{N}}$ and ${\mathbf{b}}$, respectively. The plane that is determined by ${\mathbf{N}}$ and ${\mathbf{b}}$ has an intersection curve R with the surface of budding vesicle in the opposite direction of ${\bf{N}}$. Take a point $\mathrm{Q}'$ in the opposite direction of ${\bf{N}}$ such that the distance between $\mathrm{Q}$ and $\mathrm{Q}'$ is $\epsilon$. $\mathrm{P}_1$ and $\mathrm{P}_2$ are the intersection points of the curve
R and the line that goes through point $\mathrm{Q}'$ and is parallel to ${\mathbf{b}}$. $H^\mathrm{I}_{\epsilon}$ and $H^\mathrm{II}_{\epsilon}$ in the general neck condition~(\ref{eq-neckcd}) represent the mean curvatures of membrane surface at points $\mathrm{P}_1$ and $\mathrm{P}_2$, respectively.
For the points in the vicinity of the neck, $\epsilon$ should be much smaller than the scale of the vesicle, i.e., $\epsilon\ll l_v$. Our further study will reveal that the general neck condition (\ref{eq-neckcd}) holds when
\begin{equation}\label{eq-epsilon}
l_n\ll\epsilon\ll l_i
\end{equation}
regardless of symmetry of the vesicle.

The rest of this paper is organized as follows. In Sec.~\ref{sec2}, the shape equation and linking conditions of two-domain vesicles are derived. In Sec.~\ref{sec3}, we discuss the neck condition (\ref{eq-neckcd}) in axisymmetric situation. The picture in the axisymmetric situation is relatively intuitive and the derivation is more accessible. The main findings
also offer hints to the subsequent  proof in Sec.~\ref{sec4} for the more general case
without any symmetry assumption. A brief summary is given in Sec.~\ref{sec5} and essential technical details are provided in the Appendixes at the end of this paper.

\section{Shape equation and Linking conditions of two-domain vesicles}\label{sec2}

Since the lateral dimensions of lipid vesicles are much larger than their thickness, they may be
effectively modeled as two-dimensional surfaces which are locally characterized by the mean curvature and Gaussian curvature.
Three kinds of elastic models of lipid bilayers have been proposed in the literature to analyze the shape of vesicles: the spontaneous-curvature model where a parameter $c_0$ (so called spontaneous-curvature) was introduced to reflect the asymmetric factors between the two leaves of the bilayer~\cite{Helfrich73}; the bilayer-couple model where the area of each monolayer of the bilayer was fixed~\cite{bmodel89,bilayer-model83}; the area difference elasticity model where the energy cost due the change of the area difference between the two leaves of the bilayer was included~\cite{ADEmodel1}. In this paper, we take the spontaneous-curvature model to study budding lipid vesicles following the work of J\"{u}licher and Lipowsky~\cite{LipowskyPRE96}.

\begin{figure}[htp!]
\centerline{\includegraphics[width=7cm]{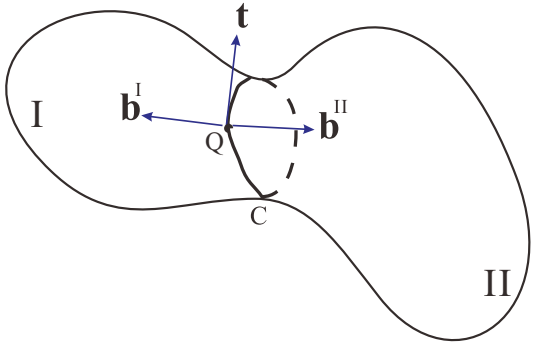}}\caption{\label{fig1-lipdm}A vesicle with two-phase domains.}
\end{figure}

Let us consider a vesicle with two-phase domains (domain I and domain II) shown in Fig.~\ref{fig1-lipdm}.
The separation boundary curve $\mathrm{C}$ is parameterized by arc length $s$. At any point $\mathrm{Q}$ on curve $\mathrm{C}$, denote $\mathbf{t}$ as the tangent vector of curve $\mathrm{C}$ at point $\mathrm{Q}$. We take two vectors $\mathbf{b}^\mathrm{I}$ and $\mathbf{b}^\mathrm{II}$ in the tangent plane of membrane surface at point $\mathrm{Q}$ with $\mathbf{b}^\alpha$ ($\alpha=\mathrm{I},\mathrm{II}$) being perpendicular to $\mathbf{t}$ and pointing to the side of domain $\alpha$. We assume that the surface is smooth enough such that $\mathbf{b}^\mathrm{II}=-\mathbf{b}^\mathrm{I}$ at any point $\mathrm{Q}$ on curve $\mathrm{C}$.

The free energy of a vesicle may be expressed as~\cite{LipowskyPRE96}:
\begin{eqnarray}\label{eq-free energy 2phase}
F&=&\frac{k_c^\mathrm{I}}{2}\int (2H^\mathrm{I}+c_0^\mathrm{I})^2 \mathrm{d}A^\mathrm{I}+\frac{k_c^\mathrm{II}}{2}\int (2H^\mathrm{II}+c_0^\mathrm{II})^2 \mathrm{d}A^\mathrm{II}\nonumber\\
&&+\lambda^\mathrm{I} A^\mathrm{I}+\lambda^\mathrm{II} A^\mathrm{II}+\gamma\oint \mathrm{d}s+pV.\label{eq-fE}
\end{eqnarray}
The first two terms represent the bending energy of both lipid domains with vanishing Gaussian bending modulus.
$H^{\alpha}$ ($\alpha=\mathrm{I},\mathrm{II}$) in the above equation represents the mean curvature of a point in domain $\alpha$. Note that the sign of mean curvatures here is opposite to those defined by J\"{u}licher and Lipowsky. $k_c^{\alpha}$, $c_0^{\alpha}$, $\lambda^{\alpha}$, $A^\alpha$, $p$ and $V$ represent the bending modulus, the spontaneous curvature, the surface tension, the surface area of domain $\alpha$ ($\alpha=\mathrm{I},\mathrm{II}$), the osmotic pressure and the volume of the whole lipid vesicle, respectively.
$\gamma$ is the line tension of the separation boundary. It is worth noting that when the Gaussian bending moduli of two domains take the same value, the free energy of a vesicle may always be expressed as Eq.~(\ref{eq-free energy 2phase}) according to Gauss-Bonnet theorem. The only difference is an insignificant constant.

The first order variation of free energy functional (\ref{eq-free energy 2phase}) can be calculated following the procedure proposed in Refs.~\cite{TuJPA04,TuJCTN08} and the shape equation valid  in  the domain $\alpha$ ($\alpha=\mathrm{I},\mathrm{II}$) can be derived:
\begin{eqnarray}\label{eq-shape}
&&k_{c}^\alpha(2H^\alpha+c_{0}^\alpha)[2(H^{\alpha})^2-c_{0}^\alpha H^\alpha-2K^\alpha]\nonumber\\
&+&k_{c}^\alpha\nabla^{2}(2H^\alpha)-2\lambda^\alpha H^\alpha+p=0.
\end{eqnarray}
Simultaneously, we can derive three linking conditions that are satisfied on the separation boundary C as well:
\begin{equation}\label{eq-bc1}
\left.k_{c}^{\mathrm{I}}(2H^{\mathrm{I}}+c_{0}^{\mathrm{I}})\right|_\mathrm{C}=\left.k_{c}^{\mathrm{II}%
}(2H^{\mathrm{II}}+c_{0}^{\mathrm{II}})\right|_\mathrm{C},
\end{equation}
\begin{equation}\label{eq-bc2}
\left.\frac{\partial\left[k_{c}^{\mathrm{I}}(2H^{\mathrm{I}}+c_{0}^{\mathrm{I}%
})\right]  }{\partial\mathbf{b}^{\mathrm{I}}}\right|_\mathrm{C}+\left.\frac{\partial\left[
k_{c}^{\mathrm{II}}(2H^{\mathrm{II}}+c_{0}^{\mathrm{II}})\right]  }%
{\partial\mathbf{b}^{\mathrm{II}}}\right|_\mathrm{C}=\gamma\kappa_{n},
\end{equation}
\begin{equation}\label{eq-bc3}
\frac{k_{c}^{\mathrm{I}}}{2}[4(H^{\mathrm{I}})^2-(c_{0}^{\mathrm{I}})^2]|_\mathrm{C}-\frac
{k_{c}^{\mathrm{II}}}{2}[4(H^{\mathrm{II}})^2-(c_{0}^{\mathrm{II}})^2]|_\mathrm{C}=\lambda
^{\mathrm{I}}-\lambda^{\mathrm{II}}+\gamma\kappa_{g},
\end{equation}
where $\kappa_n$ and $\kappa_g$ are the normal curvature and geodesic curvature of curve $\mathrm{C}$, respectively.
Directional derivatives $\mathbf{b}^{\mathrm{I}}$ and $\mathbf{b}^{\mathrm{II}}$ equals to an inner product between corresponding unit vector and the gradient, respectively.
The above three linking conditions are related to the balances of force and moment on the separation curve (detailed derivation and explanation of their physical meanings as well as the general derivations for Eqs.(\ref{eq-shape})-(\ref{eq-bc3})
are available in the Supplemental Material~\cite{sm}). It is easy to see that Eq.~(\ref{eq-shape}) is just the shape equation of lipid vesicles obtained in Refs.~\cite{oyPRL87,oyPRA89}. In addition, the above linking conditions (\ref{eq-bc1})-(\ref{eq-bc3}) degenerate to the boundary conditions of an open lipid membrane \cite{TuOY2003} if all elastic constants for domain II vanish.

\section{Proof in Axisymmetric situation}\label{sec3}

To get an intuitive picture
 and to make the derivation more accessible, we first investigate the neck condition in axisymmetric situation.

\begin{figure}[htp!]
\centerline{\includegraphics[width=7cm]{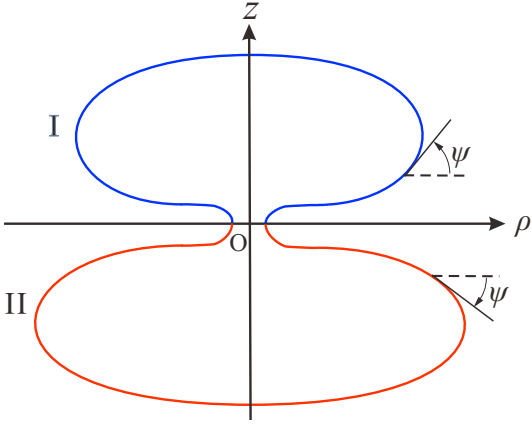}}\caption{\label{fig1-axineck}Contour line of an axisymmetric vesicle.}
\end{figure}

An axisymmetric vesicle can be generated by its contour line which is represented by $z=z(\rho)$ with $\rho$ being revolution radius.
As shown in Fig.~\ref{fig1-axineck}, the surface may be parameterized as
\begin{eqnarray}
x=\rho\cos\phi \enskip,\enskip y=\rho\sin\phi \enskip,\enskip z=\int\tan\psi(\rho)\mathrm{d}\rho
\end{eqnarray}
where $\phi$ is the azimuth angle in cylindrical coordinate. $\psi$ is the angle between the tangent of the contour line and the horizontal.

The mean curvature and Gaussian curvature can be derived as follows:
\begin{equation}\label{eq-H&G symmetric}
2H=-\left[\frac{\sin\psi}{\rho}+\frac{\mathrm{d}\left(\sin\psi\right)}{\mathrm{d}\rho}\right],\quad K=\frac{\sin\psi}{\rho}\frac{\mathrm{d}\sin\psi}{\mathrm{d}\rho}
\end{equation}
Substituting the above equations into general shape equation (\ref{eq-shape}), we obtain:
\begin{eqnarray}\label{eq-SE ax1}
&&\cos ^{3}\psi \frac{\mathrm{d} ^{3}\psi }{\mathrm{d} {\rho }^{3}}-4\sin \psi \cos
^{2}\psi \frac{\mathrm{d} \psi }{\mathrm{d} {\rho }}\frac{\mathrm{d} ^{2}\psi }{\mathrm{d}
{\rho }^{2}}+\frac{2\cos ^{3}\psi }{\rho }\frac{\mathrm{d} ^{2}\psi }{\mathrm{d} {\rho }^{2}}\nonumber\\
&&+\cos \psi \left( \sin ^{2}\psi -\frac{\cos ^{2}\psi }{2}\right) \left(
\frac{\mathrm{d} \psi }{\mathrm{d} {\rho }}\right) ^{3}-\frac{7\sin \psi \cos ^{2}\psi
}{2\rho }\left( \frac{\mathrm{d} \psi }{\mathrm{d} {\rho }}\right) ^{2} \nonumber\\
&&-\left[ \tilde{\lambda }-\frac{2c_{0}\sin \psi }{\rho }-\frac{\left(
\sin ^{2}\psi -2\cos ^{2}\psi \right) }{2\rho ^{2}}\right] \cos \psi \frac{%
\mathrm{d} \psi }{\mathrm{d} {\rho }}\nonumber\\
&&+\frac{\left( 1+\cos ^{2}\psi \right) \sin \psi }{%
2\rho ^{3}}-\frac{\tilde{\lambda }\sin \psi }{\rho }=\tilde{p}
\end{eqnarray}
where $\tilde{\lambda }={\lambda }/{k_{c}}+{c_{0}^{2}}/{2}$ and $\tilde{p}={p}/{k_c}$. The above equation is in fact identical to the axisymmetric shape equation obtained by Hu and Ou-Yang~\cite{HuPRE93}. Following Ref.~\cite{ZhengPRE93}, we can transform it into a second-order differential equation
\begin{eqnarray}\label{eq-first integral1}
\eta&=&\frac{\rho \sin \psi \cos
^{2}\psi }{2}\left( \frac{\mathrm{d}\psi }{\mathrm{d}{\rho } }\right) ^{2}-\rho \cos ^{3}\psi \frac{\mathrm{d}^{2}\psi }{\mathrm{d}{\rho } ^{2}}\nonumber\\
&&-\cos ^{3}\psi \frac{\mathrm{d}\psi }{\mathrm{d}{\rho } }+\frac{\rho \sin \psi
}{2}\left( \frac{\sin \psi }{\rho }-c_{0}\right) ^{2}\nonumber\\
&&+\rho \left( \tilde{\lambda }-\frac{c_{0}^{2}}{2}\right) \sin \psi+\frac{\sin \psi \cos ^{2}\psi }{\rho }+\frac{\tilde{p}\rho^2}{2}
\end{eqnarray}
with an integral constant $\eta$ (the so-called  first integral).

A budding vesicle can be intuitively regarded as two open lipid vesicles docking together with the same boundary curve. We assume that the separation boundary happens to be the neck. This assumption is reasonable when the Gaussian bending modulus of lipid bilayer is omitted~\cite{LipowskyPRE96}. The neck is a circle with radius  $1/\kappa$ where $\kappa$ is the curvature of the neck curve. For the limit shape, $\kappa$ is infinity since the neck is infinitesimal. Membrane in the vicinity of the neck is highly singular where two principal curvatures with opposite signs are on the order of magnitude much larger than $1/l_v$. The Gaussian curvature is singular while the mean curvature may be finite in the vicinity of the neck. In this paper we only consider the situation that the mean curvature is finite, otherwise the general neck condition would not be true. Next we will analyze the local behavior of the membrane segments adjacent to the neck.

Introducing an auxiliary function
\begin{eqnarray}\label{eq-g define}
\Phi({\rho })=-(2H+c_{0})=\frac{\sin \psi }{\rho }+\frac{\mathrm{d}(\sin \psi
) }{\mathrm{d}{\rho } }-c_{0}.
\end{eqnarray}
and substituting it into Eq.(\ref{eq-first integral1}), we may achieve:
\begin{eqnarray}\label{eq-eta20}
\eta=&&\frac{\sin ^{3}\psi }{2\rho }-\frac{\rho \sin \psi }{2}\left( \Phi+c_{0}-\frac{%
\sin \psi }{\rho }\right) ^{2}-c_{0}\sin ^{2}\psi\nonumber\\
&&-\rho \left( 1-\sin ^{2}\psi \right) \frac{%
\mathrm{d} \Phi}{\mathrm{d} {\rho }}+\tilde{\lambda }\rho \sin
\psi+\frac{\tilde{p}\rho^2}{2}.
\end{eqnarray}
By considering the natural boundary condition that $\sin\psi=1$ at the neck, we obtain the integral constant
\begin{eqnarray}\label{eq-eta}
\eta =\left[ \tilde{\lambda }-\frac{\left(
\Phi_{0}+c_{0}\right) ^{2}}{2}\right] \frac{1}{\kappa}+\Phi_{0}+\frac{\tilde{p}}{2\kappa^2}
\end{eqnarray}
with $\Phi_0=\Phi({1}/{\kappa})$.

Considering the definition of $\Phi$, i.e., Eq.~(\ref{eq-g define}), we may solve
\begin{eqnarray}\label{eq-sinwhole}
\sin \psi &=&\frac{1}{\rho\kappa }+\frac{c_{0}\left( \rho ^{2}-1/\kappa^{2}\right) }{2\rho }+\frac{1}{\rho }\int_{\frac{1}{\kappa}}^{\rho }\rho\Phi \mathrm{d}\rho\nonumber\\
&=&\frac{1}{\rho \kappa }+\frac{c_{0}u}{2}\left[ \frac{u\kappa+2 }{%
u\kappa+1 }+\frac{2}{c_{0}u}\int_{0}^{u}\frac{u'\kappa +1
}{ u\kappa+1  }\Phi\mathrm{d}u' \right].
\end{eqnarray}
When writing the second term on the second line of the above equation, we have changed the variable $\rho$ to $u=\rho-1/\kappa$. This term is of the same order of $c_0u$ for the bounded function $\Phi$, thus $\sin\psi$ may be further reduced to a concise form:
\begin{eqnarray}\label{eq-sin}
\sin \psi =\frac{1}{\rho\kappa }+O(c_0u).
\end{eqnarray}

If we are only concerned with the local shape of the membrane in the scale much smaller than the intermediate length scale, i.e.,
\begin{eqnarray}\label{eq-uscope1}
u=\rho-{1}/{\kappa}\ll l_i\le\sqrt{1/{c_0\kappa}},
\end{eqnarray}
we can readily see $c_0u\ll{{1}/{\rho\kappa}}$ and $\sin \psi \approx {1}/{\rho\kappa }$ because $1/\kappa\ll 1/c_0$. Then ${\tilde{p}\rho^2}/{2}$ may be neglected since it is much smaller than $\tilde{\lambda }\rho \sin
\psi$ in Eq.~(\ref{eq-eta20}) with the consideration of $\rho\ll l_i\le \sqrt{\lambda/(p\kappa)}$. It should be noticed that $c_0$ in (\ref{eq-uscope1}) refers to the larger
one of $c_0^{\mathrm{I}}$ and $c_0^{\mathrm{II}}$.
Considering this point, substituting Eqs.~(\ref{eq-eta}) and (\ref{eq-sin}) into Eq.~(\ref{eq-eta20}), we obtain
\begin{eqnarray}\label{eq-g full}
\frac{\Phi}{\rho ^{2}\kappa ^{2}}&-&\frac{\rho ^{2}-\left( 1/\kappa \right) ^{2}%
}{\rho }\frac{\mathrm{d}\Phi}{\mathrm{d}\rho }-\Phi_{0}\nonumber\\
&-&\frac{\left( \Phi^{2}-\Phi_{0}^{2}\right) }{2\kappa }-\frac{c_{0}\left( \Phi-\Phi_{0}\right) }{\kappa }=0.
\end{eqnarray}
From Eq.~(\ref{eq-uscope1}) we have $c_0/\kappa\ll 1/\rho^2\kappa^2$, which implies that the last two terms of the above equation may be neglected.
Then the above equation is transformed into the following concise form:
\begin{equation}\label{eq-g approximatel}
\frac{\Phi}{\rho ^{2}\kappa ^{2}}-\frac{\rho ^{2}-\left( 1/\kappa \right) ^{2}%
}{\rho }\frac{\mathrm{d}\Phi}{\mathrm{d}\rho }-\Phi_{0}=0.
\end{equation}
The full solution to the above equation is
\begin{eqnarray}
\Phi&= &\Phi_{0}\left[ 1-\sqrt{1-\frac{1}{\rho ^{2}{\kappa^2}}}\ln \left(
\rho\kappa+\sqrt{{\rho ^{2}}{\kappa^{2}}-1}\right) \right]\nonumber\\
&&+B\sqrt{1-\frac{1}{\rho ^{2}\kappa^2}},
\end{eqnarray}
where $B$ is a constant.
We find that $\Phi_0$ should be 0, otherwise the local free energy for the membrane segments adjacent to the neck is quite large, which is unfavourable for minimizing the free energy. A detailed discussion can be found in Appendix~\ref{secappE}. Therefore the physically acceptable solution is
\begin{eqnarray}\label{eq-g}
\Phi(\rho)&=&B\sqrt{1-\frac{1}{\rho ^{2}\kappa^2}}
\end{eqnarray}
when $\rho\ll l_i$.

Note that the above equations (\ref{eq-SE ax1})--(\ref{eq-g}) hold for both domains. The parameters $k_c$, $c_0$, $\lambda$, $\eta$ and $B$ correspond to $k_c^{\mathrm{I}}$, $c_0^{\mathrm{I}}$, $\lambda^{\mathrm{I}}$, $\eta^{\mathrm{I}}$ and $B^{\mathrm{I}}$ for domain ${\mathrm{I}}$. The same notation is applicable to domain ${\mathrm{II}}$.

Next we turn to the linking conditions (\ref{eq-bc1})-(\ref{eq-bc3}), which may be expressed as
\begin{equation}\label{eq-axbc1}
\left.k_{c}^{\mathrm{I}}\Phi^{\mathrm{I}}\right|_{\rho=1/\kappa}=\left.k_{c}^{\mathrm{II}}\Phi^{\mathrm{II}}\right|_{\rho=1/\kappa},
\end{equation}
\begin{equation}\label{eq-axbc2}
\left.k_{c}^{\mathrm{I}}\frac{\mathrm{d} \Phi^{\mathrm{I}}}{\mathrm{d} \rho }\cos \psi\right|_{\rho=1/\kappa} +\left.k_{c}^{\mathrm{II}}\frac{%
\mathrm{d} \Phi^{\mathrm{II}}}{\mathrm{d} \rho }\cos \psi\right|_{\rho=1/\kappa}=\gamma \kappa,
\end{equation}
\begin{equation}\label{eq-axbc3}
\left.k_{c}^{\mathrm{I}}\Phi^{\mathrm{I}}(\Phi^{\mathrm{I}}-2c_0^{\mathrm{I}})\right|_{\rho=1/\kappa}
-\left.k_{c}^{\mathrm{II}}\Phi^{\mathrm{II}}(\Phi^{\mathrm{II}}-2c_0^{\mathrm{II}})\right|_{\rho=1/\kappa}
=2(\lambda^{\mathrm{I}}-\lambda^{\mathrm{II}})
\end{equation}
in axisymmetric situation, respectively.
The solution (\ref{eq-g}) automatically satisfies linking condition (\ref{eq-axbc1}).
Substituting ($\ref{eq-g}$) into (\ref{eq-axbc2}), we have
\begin{eqnarray}\label{eq-kB}
k_{c}^{\mathrm{I}}B^{\mathrm{I}}+k_{c}^{\mathrm{II}}B^{\mathrm{II}}=\gamma
\end{eqnarray}

Now let us turn our attention back to the neck condition (\ref{eq-neckcd}). With the consideration of
Eqs.~(\ref{eq-g define}), (\ref{eq-g}), (\ref{eq-kB}), and the geometric definition of $\epsilon$ shown in
Fig.\ref{fig0-epsilon}, we may obtain:
\begin{eqnarray}\label{eq-neckcd2}
&&k_{c}^{\mathrm{I}}\left( 2H_{\epsilon }^{\mathrm{I}}+c_{0}^{\mathrm{I}}\right) +k_{c}^{\mathrm{II}}\left(
2H_{\epsilon }^{\mathrm{II}}+c_{0}^{\mathrm{II}}\right) \nonumber\\
=&&k_{c}^{\mathrm{I}}\left( \left. -\Phi^{\mathrm{I}}\right|_{\rho=\epsilon+1/\kappa}\right)+k_{c}^{\mathrm{II}}\left( \left.-\Phi^{\mathrm{II}}\right|_{\rho=\epsilon+1/\kappa}\right)\nonumber\\
=&&-\gamma \sqrt{1-{1}/{{(\epsilon\kappa+1)}^2}}.
\end{eqnarray}
We find that when $\epsilon\gg{1}/{\kappa}$, Eq.~(\ref{eq-neckcd2}) leads to the neck condition (\ref{eq-neckcd}). Since (\ref{eq-g}) holds for $\rho\ll l_i$, $\epsilon$ should be also much smaller than $l_i$.
Thus the neck condition is true in the region adjacent to the neck as described in (\ref{eq-epsilon}).

\section{general proof}\label{sec4}
Non-axisymmetric budding as a common pattern has been experimentally observed when studying the mitosis process of a budding yeast~\cite{StraightCB96} and in the budding process of a binary vesicle composed of 1,2-dipalmitoyl-$sn$-glycero-3-phosphocholine (DPPC) and 1,2-dilauroyl-$sn$-glycero-3-phosphoethanolamine(DLPE)~\cite{Sakuma2015}. Recent numerical work~\cite{Sauerjcp2017} suggests that non-axisymmetric budding is not only observed, but also preferred over axisymmetric one.
In this section we will give a general proof without the axisymmetric assumption.

\begin{figure}[htp!]
\centerline{\includegraphics[width=7cm]{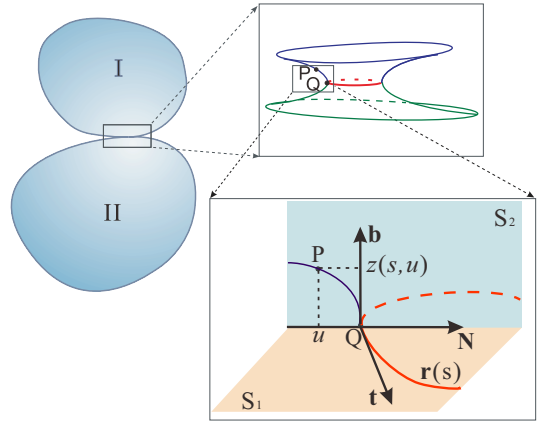}}\caption{\label{fig1-generalneck}Local surface in the vicinity of the neck.}
\end{figure}

The discussion in the above section reveals that only the local behavior plays a role in the proof of the neck condition. On account of this point, we first parametrize the local surface adjacent to the neck.
The neck is described as ${\bf{r}}(s)$ with $s$ being the arclength parameter, which does not have to be a planar curve.
As shown in Fig.~\ref{fig1-generalneck}, at any point $\mathrm{Q}$ on the neck ${\bf{r}}(s)$, the tangent vector, the normal vector and the binormal vector are denoted as ${\mathbf{t}}$, ${\mathbf{N}}$ and ${\mathbf{b}}$, respectively. $\mathrm{S}_1$ is the plane determined by ${\bf{t}}$ and ${\bf{N}}$, while $\mathrm{S}_2$ is determined by ${\bf{N}}$ and ${\bf{b}}$. Any point $\mathrm{P}$ on the intersection curve between the plane $\mathrm{S}_2$ and the membrane surface may be expressed as
a vector
\begin{eqnarray}\label{eq-y}
{\bf{Y}}(s,u)={\bf{r}}(s)-u{\bf{N}}+z\left(
s,u\right) {\bf{b}}
\end{eqnarray}
where the parameter $u$ represents the distance between the projection of P on plane $\mathrm{S}_1$ and point $\mathrm{Q}$. $z=z(s,u)$ is the distance from P to plane $\mathrm{S}_1$.

The above equation (\ref{eq-y}) is actually a local parametrization of the membrane surface in the neck region. The shape of membrane in the vicinity of the neck is determined not only by $\mathbf{r}(s)$ but also by $z(s,u)$. The membrane surface adjacent to the neck is highly singular where two principal curvatures with opposite signs are on the order of magnitude much larger than $1/l_v$. One principal curvature is of the order of $\kappa(s)$ which is the curvature of the neck at point $\mathrm{Q}$. The other one is on the order of the curvature of the contour curve $z=z(s,u)$ for given $s$, which can be easily calculated as $-{z_{uu}}/{\left( 1+z_{u}^{2}\right) ^{\frac{3}{2}}}$ where $z_u$ and $z_{uu}$ represent the first and the second derivatives of $z$ with respect to $u$, respectively. Since the mean curvature, that is the sum of both principal curvatures,  is finite, we see that the two principal curvatures should be on the same order, which implies $ z_{uu}\sim\kappa z_{u}^{3}$.
The latter principal curvature mentioned above is a large magnitude relative to $1/l_v$, which implies that $z$ varies quickly with respect to $u$ in the neck region.
On the other hand, from the geometric point of view the shape of the contour curve $z'=z(s+\Delta s,u)$ departs slightly from that of $z=z(s,u)$ for small $\Delta s$.
Therefore we make the following reasonable assumption on the membrane adjacent to the neck: $z(s,u)$ is the fast variable with respect to $u$ but a slow variable with respect to $s$. Besides, we also assume that $\kappa(s)$ varies not too quickly with respect to $s$, though the magnitude of $\kappa(s)$ itself is much larger than $1/l_v$.

Under the above assumptions we may derive the leading order of the mean curvature and the Gaussian curvature as below:
\begin{eqnarray*}%\label{eq-H&G without symm}
2H&=&-\frac{z_{u}}{\left( u+\frac{1}{\kappa }\right) \sqrt{1+z_{u}^{2}}}-\frac{%
z_{uu}}{\left( 1+z_{u}^{2}\right) ^{\frac{3}{2}}},\nonumber\\
K&=&\frac{z_{uu}z_{u}}{%
\left( 1+z_{u}^{2}\right) ^{2}\left( u+\frac{1}{\kappa }\right) }.
\end{eqnarray*}
Detailed derivations can be found in Appendix~\ref{secappF}. By
introducing two new variables $\psi\equiv\arctan {z_u}$ and $\rho\equiv u+1/\kappa(s)$, the above two curvatures can be expressed as
\begin{equation}\label{eq-2H-general}
2H= -\frac{\sin \psi }{\rho }-\frac{\partial  \sin \psi
 }{\partial u}%
,\quad K=\frac{\sin\psi}{\rho}\frac{\partial\sin\psi}{\partial u}.
\end{equation}
We further derive
\begin{eqnarray}
\nabla ^{2}\left( 2H\right)
=\frac{\cos \psi }{\rho }\frac{\partial }{\partial u}\left[ \rho \cos \psi
\frac{\partial \left( 2H\right) }{\partial u}\right]
\end{eqnarray}
through some tedious calculations and comparison on the orders of magnitude.
Detailed derivations can be found in Appendix~\ref{secappG}.

Substituting the above two equations into shape equation (\ref{eq-shape}), we obtain
\begin{eqnarray}\label{eq-SE ge}
&&\cos ^{3}\psi \frac{\partial ^{3}\psi }{\partial u^{3}}-4\sin \psi \cos
^{2}\psi \frac{\partial \psi }{\partial u}\frac{\partial ^{2}\psi }{\partial
u^{2}}+\frac{2\cos ^{3}\psi }{\rho }\frac{\partial ^{2}\psi }{\partial u^{2}}\nonumber\\%
&&+\cos \psi \left( \sin ^{2}\psi -\frac{\cos ^{2}\psi }{2}\right) \left(
\frac{\partial \psi }{\partial u}\right) ^{3}-\frac{7\sin \psi \cos ^{2}\psi
}{2\rho }\left( \frac{\partial \psi }{\partial u}\right) ^{2} \nonumber\\
&&-\left[ \tilde{\lambda }-\frac{2c_{0}\sin \psi }{\rho }-\frac{\left(
\sin ^{2}\psi -2\cos ^{2}\psi \right) }{2\rho ^{2}}\right] \cos \psi \frac{%
\partial \psi }{\partial u}\nonumber\\
&&+\frac{\left( 1+\cos ^{2}\psi \right) \sin \psi }{%
2\rho ^{3}}-\tilde{\lambda }\frac{\sin \psi }{\rho }=\tilde{p}
\end{eqnarray}
where $\tilde{\lambda }={\lambda }/{k_{c}}+{c_{0}^{2}}/{2}$ and $\tilde{p}={p}/{k_c}$.
The curvature $\kappa(s)$ is a constant in axisymmetric situation, which implies $\partial/\partial u=\mathrm{d}/\mathrm{d}\rho$,
thus the above equation (\ref{eq-SE ge}) degenerates to (\ref{eq-SE ax1}).
Similarly, equation (\ref{eq-SE ge}) may also be transformed into a second-order equation:
\begin{eqnarray}\label{eq-first integral2}
\eta(s)=&&\frac{\rho \sin \psi \cos
^{2}\psi }{2}\left( \frac{{\partial}\psi }{{\partial}u }\right) ^{2}-\rho \cos ^{3}\psi \frac{{\partial}^{2}\psi }{{\partial}u^{2}}\nonumber\\
&&-\cos ^{3}\psi \frac{{\partial}\psi }{{\partial}u }+\frac{\rho \sin \psi
}{2}\left( \frac{\sin \psi }{\rho }-c_{0}\right) ^{2}\nonumber\\
&&+\rho \left( \tilde{\lambda }-\frac{c_{0}^{2}}{2}\right) \sin \psi+\frac{\sin \psi \cos ^{2}\psi }{\rho }+\frac{\tilde{p}\rho^2}{2}.
\end{eqnarray}
The above equation degenerates to (\ref{eq-first integral1}) with $\eta(s)$ being a constant in axisymmetric situation.

Now, let us introduce an auxiliary function
\begin{eqnarray} \label{eq-Psi}
\Psi(s,u)&=&-(2H+c_0)\nonumber\\
&=&\frac{\sin \psi }{u+1/\kappa(s) }+\frac{\partial  \sin \psi}{\partial u}-c_0.
\end{eqnarray}
Substituting (\ref{eq-Psi}) into (\ref{eq-first integral2}), by analogy with similar discussion in axisymmetric situation, we may obtain
\begin{eqnarray}\label{eq-Psi2}
\frac{\Psi}{\left[1+u\kappa(s)\right]^2}-\frac{\left[1+u\kappa(s)\right]^2-1}{\left[1+u\kappa(s)\right]\kappa(s) }\frac{\partial \Psi}{\partial u}-\Psi_{0}=0
\end{eqnarray}
with $\Phi_0=\Psi(s,0)$ when $u\ll l_i$.
A physically acceptable solution for (\ref{eq-Psi2}) is
\begin{eqnarray}\label{eq-solvedPsi}
\Psi( s,u)  &=&B\sqrt{1-\frac{1}{\left[1+u\kappa(s)\right]^2}},
\end{eqnarray}
where $B$ is a constant.

Note that the above equations (\ref{eq-SE ge})-(\ref{eq-solvedPsi}) hold for both domains. The parameters $k_c$, $c_0$, $\lambda$ and $B$ correspond to $k_c^{\mathrm{I}}$, $c_0^{\mathrm{I}}$, $\lambda^{\mathrm{I}}$ and $B^{\mathrm{I}}$ for domain ${\mathrm{I}}$. The same notation is applicable for domain ${\mathrm{II}}$.

Then we turn to the linking conditions (\ref{eq-bc1})-(\ref{eq-bc3}), which may be expressed as
\begin{equation}\label{eq-gebc1}
\left.k_{c}^{\mathrm{I}}\Psi^{\mathrm{I}}\right|_{u=0}=\left.k_{c}^{\mathrm{II}}\Psi^{\mathrm{II}}\right|_{u=0},
\end{equation}
\begin{equation}\label{eq-gebc2}
\left.k_{c}^{\mathrm{I}}\frac{\partial \Psi^{\mathrm{I}}}{\partial u }\cos \psi\right|_{u=0} +\left.k_{c}^{\mathrm{II}}\frac{%
\partial \Psi^{\mathrm{II}}}{\partial u }\cos \psi\right|_{u=0} =\gamma \kappa(s)
\end{equation}
\begin{equation}\label{eq-gebc3}
\left.k_{c}^{\mathrm{I}}\Psi^{\mathrm{I}}(\Psi^{\mathrm{I}}-2c_0^{\mathrm{I}})\right|_{u=0}
-\left.k_{c}^{\mathrm{II}}\Psi^{\mathrm{II}}(\Psi^{\mathrm{II}}-2c_0^{\mathrm{II}})\right|_{u=0}
=2(\lambda^{\mathrm{I}}-\lambda^{\mathrm{II}})
\end{equation}
 respectively.
The above solution (\ref{eq-solvedPsi}) automatically satisfies linking condition (\ref{eq-gebc1}).
Substituting ($\ref{eq-solvedPsi}$) into (\ref{eq-gebc2}), we have
\begin{eqnarray}\label{eq-kB2}
k_{c}^{\mathrm{I}}B^{\mathrm{I}}+k_{c}^{\mathrm{II}}B^{\mathrm{II}}=\gamma
\end{eqnarray}

Now we turn our attention back to the general neck condition (\ref{eq-neckcd}). With Eqs.~(\ref{eq-Psi}), (\ref{eq-solvedPsi}), (\ref{eq-kB2}), and the geometric definition of $\epsilon$ shown in Fig.\ref{fig0-epsilon}, we may obtain
\begin{eqnarray}\label{eq-neckcd3}
&&k_{c}^{\mathrm{I}}\left( 2H_{\epsilon }^{\mathrm{I}}+c_{0}^{\mathrm{I}}\right) +k_{c}^{\mathrm{II}}\left(
2H_{\epsilon }^{\mathrm{II}}+c_{0}^{\mathrm{II}}\right) \nonumber\\
=&&k_{c}^{\mathrm{I}}\left( \left. -\Psi^{\mathrm{I}}\right|_{u=\epsilon}\right)+k_{c}^{\mathrm{II}}\left( \left.-\Psi^{\mathrm{II}}\right|_{u=\epsilon}\right)\nonumber\\
=&&-\gamma \sqrt{1-\frac{1}{\left[1+\epsilon\kappa(s)\right]^2}}.
\end{eqnarray}
We find that when $\epsilon\gg1/\kappa(s)$, the above equation returns the neck condition (\ref{eq-neckcd}). Since the solution (\ref{eq-solvedPsi}) holds for $u\ll l_i$, $\epsilon$ should be also much smaller than $l_i$. Thus the neck condition is true in the region adjacent to the neck as described in (\ref{eq-epsilon}).

\section{conclusion}\label{sec5}

In the above discussions we have refined and proved the conjecture on the general neck condition (\ref{eq-neckcd}) proposed by J\"{u}licher and Lipowsky. Our study reveals that the mean curvature of the membrane segments adjacent to the neck satisfies the general neck condition for the limit shape of a budding vesicle when the length scale of the membrane segments is much larger than the characteristic size of the neck but still much smaller than the characteristic size of the vesicle.
In the more general proof, we did not introduce any axis-symmetry assumption or special trial configuration, which implies that the elegant neck condition (\ref{eq-neckcd}) is indeed broadly applicable. From the derivations given in our proof,
we see that the local neck condition for the limit shape of a budding vesicle is unaffected by the global shape of the vesicle. In other words, the relationship between the mean curvature of membrane segments adjacent to the neck and the spontaneous curvature of the membrane is determined by the line tension of the separation curve, which would not be affected by the specific morphology of two daughter vesicles.

The general neck condition (\ref{eq-neckcd}) degenerates to a more concise form for a budding vesicle with uniform phase. Since $k_c^\mathrm{I}=k_c^\mathrm{II}\equiv k_c$ and $c_0^\mathrm{I}=c_0^\mathrm{II}\equiv c_0$ in the uniform phase, from (\ref{eq-neckcd}) we obtain
\begin{equation}\label{eq-neckcduniform}
H^\mathrm{I}_{\epsilon}+H^\mathrm{II}_{\epsilon}+c_0+\gamma/2k_c=0.
\end{equation}
If we consider a special limit shape consisting of two spheres connected by an infinitesimal neck,
the mean curvatures of two spheres can be expressed as $H^\mathrm{I}_{\epsilon}=-1/R^\mathrm{I}$ and $H^\mathrm{II}_{\epsilon}=-1/R^\mathrm{II}$, where $R^\mathrm{I}$ and $R^\mathrm{II}$ represent the radii of two spheres, respectively. Then the above equation is transformed into
\begin{equation}\label{eq-neckcduniform2}
1/R^\mathrm{I}+1/R^\mathrm{II}=c_0+\gamma/2k_c,
\end{equation}
which implies that one may experimentally estimate the force generated by the contractile ring by measuring the sizes of two daughter cells.

It is worth emphasizing several notable features of our proof.
Firstly, we draw lessons from the idea of separation of fast and slow variables. When specifying the local behavior of membrane surface adjacent to the neck of a budding vesicle, we assume that $z(s,u)$ varies quickly with respect to $u$ but slowly with respect to $s$. This leads to a concise local shape equation (\ref{eq-SE ge}) which has the similar form as the axisymmetric shape equation (\ref{eq-SE ax1}). Such a consequence is consistent with our expectation that any
finite deviation from the axis-symmetry is insignificant in the region close to the singular set,
which in our discussion refers to the neck curve of the budding vesicle in the limit shape. In addition, multiscale analysis is used in the proof. We introduce three length scales including a macroscopic scale $l_v$, a microscopic scale $l_n$ and an intermediate scale $l_i$ which is macroscopically infinitesimal but microscopically infinite large. Based on the multiscale analysis we can give a quantitative definition (\ref{eq-epsilon}) of what means to be ``adjacent to the neck".

We would like to give some remarks on several open questions in the end.
The Gaussian bending energy has not been taken into account in the free energy (\ref{eq-free energy 2phase}). When the Gaussian bending moduli of two domains of a budding vesicle take on the same value, the neck condition (\ref{eq-neckcd}) still holds according to the Gauss-Bonnet theorem. However, if the Gaussian bending moduli of two domains differ, the separation boundary between two domains may not be the neck any more~\cite{LipowskyPRE96}. Thus the neck condition (\ref{eq-neckcd}) is no longer applicable. In this situation whether there exists a more general neck condition is yet unclear.
In addition, we have made a hypothesis of finite mean curvature in the vicinity of the neck to prove the conjecture of general neck condition. This hypothesis has been adopted in previous studies as well~\cite{Seifert91,LipowskyPRL93,LipowskyPRE96}.
We expect that this hypothesis can be derived from a more fundamental principle. This issue might be resolved with further physical considerations of the boundedness of the free energy and free energy density.
Furthermore, in the present work we merely consider the vesicle of external budding where two domains are located on different sides of the neck. We are not concerned with the internal budding in which the daughter vesicle is produced inside a mother vesicle.  The latter kind of budding has been observed in many cellular processes, such as endocytosis, autophagy and so on~\cite{Knorr15,Mizushima11,Canalejo15,Canalejo16}.
The general neck condition in such a situation needs to be further investigated.

\section*{Acknowledgement}
The authors are grateful to financial support from the
National Natural Science Foundation of China (Grants No. 11322543 and No. 11274046) and
US National Science Foundation (DMS-1558744).
They also thank Jaime Agudo-Canalejo, Xin Zhou, Hepeng Zhang and Pablo V\'{a}zquez-Montejo for their instructive discussions.

\appendix

\section{The solution to Equation (\ref{eq-g approximatel})}\label{secappE}

The general solution to Eq. (\ref{eq-g approximatel}) is
\begin{eqnarray}
\Phi &=& \Phi_{0}\left[ 1-\sqrt{1-\frac{1}{\rho ^{2}{\kappa^2}}}\ln \left(
\rho\kappa+\sqrt{{\rho ^{2}}{\kappa^{2}}-1}\right) \right] \nonumber\\
&+&B\sqrt{1-\frac{1}{\rho ^{2}\kappa^2}}.
\end{eqnarray}

We discuss the free energy of a ribbon with radius $\rho$ between $\xi$ to $2\xi$. The specific value of $\xi$ is taken to
satisfy  $1/\kappa\ll \xi  \ll l_i$. When $1/\kappa\ll\xi<\rho<2\xi\ll l_i$, $\Phi\approx\Phi_{0}\ln(2\rho\kappa)+B$.
The free energy of the ribbon turns out to be
\begin{eqnarray}
&&\int(2H+c_0)^2\mathrm{d}A\nonumber\\
&=&\int_0^{2\pi}\mathrm{d}\phi\int_{\xi}^{2\xi}{\Phi}^2\frac{\rho }{\sqrt{1-\frac{1}{\rho^2\kappa^2}} }\mathrm{d}\rho\nonumber\\
&\approx&2\pi\int_{\xi}^{2\xi} \left[B+\Phi_{0}\ln(2\rho\kappa)\right]^2\rho\mathrm{d}\rho \nonumber\\
&=&2\pi\left[B+\Phi_{0}\ln(2\bar{\rho}\kappa)\right]^2\bar{\rho}\xi
\end{eqnarray}
with $\xi<\bar{\rho}<2\xi$.
The last equality is due to the mean value theorem of integral form.
When $\bar{\rho}\gg 1/\kappa$, the term containing $\ln(2\bar{\rho}\kappa)$ would be a relatively large term, which is unfavourable for minimizing the free energy. Thus a reasonable choice is  $\Phi_0=0$.

\section{Derivation of mean curvature and Gaussian curvature}\label{secappF}

According to the local parametrization
\begin{eqnarray}
{\bf{Y}}(s,u)={\bf{r}}(s)-u{\bf{N}}+z\left(
s,u\right) {\bf{b}}
\end{eqnarray}
we may derive the following equations
\begin{eqnarray}
{\bf Y}_{s}(s,u)&=&\left( 1+\kappa u\right) {\bf t}%
-z\tau {\bf N}+\left( z_{s}-u\tau \right) {\bf b},\\
{\bf Y}_{u}(s,u)&=&-{\bf N}+z_{u}{\bf b},\\
{\bf Y}_{su}(s,u)&=&\kappa {\bf t}-z_{u}\tau
{\bf N}+\left( z_{su}-\tau \right) {\bf b},\\
{\bf Y}_{ss}(s,u)&=&\left( u\kappa _{s}+z\tau \kappa \right)
{\bf t}+\left( z_{ss}-z\tau ^{2}\right) {\bf b}\nonumber,\\
&&+\left( \kappa +u\kappa ^{2}+u\tau ^{2}-2z_{s}\tau \right)
{\bf N},\\
{\bf Y}_{uu}(s,u)&=&z_{uu}{\bf b}
\end{eqnarray}
by means of the Frenet formula
\begin{eqnarray}
\left[
\begin{array}{c}
{\bf t}_s \\
{\bf N}_s \\
{\bf b}_s%
\end{array}%
\right] =\left[
\begin{array}{ccc}
0 & \kappa (s) & 0 \\
-\kappa (s) & 0 & \tau (s) \\
0 & -\tau (s) & 0%
\end{array}%
\right] \left[
\begin{array}{c}
{\bf t} \\
{\bf N} \\
{\bf b}%
\end{array}%
\right].
\end{eqnarray}
Here ${\bf Y}_{s}$ and ${\bf Y}_{ss}$ represent the first and the second derivatives of ${\bf Y}$ with respect to $s$. The same notation is taken for other quantities such as ${\bf t}$, ${\bf N}$, ${\bf b}$ and $z$.

The coefficients of the first fundamental form of the surface may be deduced as:
\begin{eqnarray}
g_{11}&=&{\bf Y}_{s}\cdot {\bf Y}_{s}=\left( 1+\kappa
u\right) ^{2}+\left( z_{s}-u\tau \right) ^{2}+z^{2}\tau ^{2},\label{g11}\\
g_{12}&=&{\bf Y}_{s}\cdot {\bf Y}_{u}=z\tau +\left(
z_{s}-u\tau \right) z_{u},\label{g12}\\
g_{22}&=&{\bf Y}_{u}\cdot {\bf Y}_{u}=1+z_{u}^{2}.\label{g22}
\end{eqnarray}
The normal vector of the surface is
\begin{eqnarray}
{\bf n} &=&\frac{{\bf Y}_{s}\times {\bf Y}%
_{u}}{\left\vert {\bf Y}_{s}\times {\bf Y}_{u}%
\right\vert } \nonumber\\
&=&\frac{\left( z_{s}-u\tau -z_{u}z\tau \right) {\bf t}-\left(
1+\kappa u\right) z_{u}{\bf N}}{\sqrt{\left( z_{s}-u\tau -z_{u}z\tau \right) ^{2}+\left(
1+\kappa u\right) ^{2}\left( z_{u}^{2}+1\right) }}\nonumber\\
&&-\frac{\left( 1+\kappa u\right)
{\bf b}}{{\sqrt{\left( z_{s}-u\tau -z_{u}z\tau \right) ^{2}+\left(
1+\kappa u\right) ^{2}\left( z_{u}^{2}+1\right) }}}.
\end{eqnarray}
The coefficients of the second fundamental form may also be expressed as
\begin{eqnarray}
L_{11} &=&{\bf Y}_{ss}\cdot {\bf n} \nonumber\\
&=&-\frac{z_{u}\left( u^{2}\kappa ^{2}+u\kappa +u^{2}\tau ^{2}+z^{2}\tau
^{2}-2uz_{s}\tau \right) }{\sqrt{\left( z_{s}\frac{1}{\kappa }-u\frac{\tau }{%
\kappa }-z_{u}z\frac{\tau }{\kappa }\right) ^{2}+\left( \frac{1}{\kappa }%
+u\right) ^{2}\left( z_{u}^{2}+1\right) }} \nonumber\\
&&-\frac{z_{u}\left( zu\kappa _{s}\frac{\tau }{\kappa }+u\tau \frac{\tau }{%
\kappa }-2z_{s}\frac{\tau }{\kappa }+z\tau _{s}\frac{1}{\kappa }+uz\tau
_{s}\right) }{\sqrt{\left( z_{s}\frac{1}{\kappa }-u\frac{\tau }{\kappa }%
-z_{u}z\frac{\tau }{\kappa }\right) ^{2}+\left( \frac{1}{\kappa }+u\right)
^{2}\left( z_{u}^{2}+1\right) }} \nonumber\\
&&+\frac{\left( \frac{1}{\kappa }u\kappa _{s}+z\tau \right) \left(
z_{s}-u\tau \right) }{\sqrt{\left( z_{s}\frac{1}{\kappa }-u\frac{\tau }{%
\kappa }-z_{u}z\frac{\tau }{\kappa }\right) ^{2}+\left( \frac{1}{\kappa }%
+u\right) ^{2}\left( z_{u}^{2}+1\right) }} \nonumber\\
&&-\frac{\left( z_{ss}-z\tau ^{2}-u\tau _{s}\right) \left( \frac{1}{\kappa }%
+u\right) }{\sqrt{\left( z_{s}\frac{1}{\kappa }-u\frac{\tau }{\kappa }-z_{u}z%
\frac{\tau }{\kappa }\right) ^{2}+\left( \frac{1}{\kappa }+u\right)
^{2}\left( z_{u}^{2}+1\right) }},\\\label{l11}
L_{12} &=&{\bf Y}_{su}\cdot {\bf n} \nonumber\\
&=&\frac{z_{u}^{2}\tau \left( \frac{1}{\kappa }+u\right) -z_{u}z\tau +z_{s}%
}{\sqrt{\left( z_{s}\frac{1}{\kappa }-u\frac{\tau }{\kappa }-z_{u}z\frac{%
\tau }{\kappa }\right) ^{2}+\left( \frac{1}{\kappa }+u\right) ^{2}\left(
z_{u}^{2}+1\right) }} \nonumber\\
&&-\frac{u\tau +\left( z_{su}-\tau \right) \left( \frac{1}{\kappa }+u\right)
}{\sqrt{\left( z_{s}\frac{1}{\kappa }-u\frac{\tau }{\kappa }-z_{u}z\frac{%
\tau }{\kappa }\right) ^{2}+\left( \frac{1}{\kappa }+u\right) ^{2}\left(
z_{u}^{2}+1\right) }} ,\\\label{l12}
L_{22}&=&{\bf Y}_{uu}\cdot {\bf n} \nonumber\\
&=&\frac{-\left( \frac{1}{\kappa }+u\right) z_{uu}}{\sqrt{\left( z_{s}\frac{1%
}{\kappa }-u\frac{\tau }{\kappa }-z_{u}z\frac{\tau }{\kappa }\right)
^{2}+\left( \frac{1}{\kappa }+u\right) ^{2}\left( z_{u}^{2}+1\right) }}.\label{l22}
\end{eqnarray}

According to our assumption about the fast and slow variables, we know that $z_u\gg z_s$ and $z_{uu}\gg z_{us}$. Meanwhile though the magnitude of $\kappa(s)$ itself is much larger than $1/l_v$, $\kappa(s)$ does not vary  so quickly with respect to $s$. The torsion of neck curve is assumed to be finite, $\tau\ll \kappa$. In the vicinity of the neck, $z$ and $u$ are much smaller than the characteristic length of the vesicle.
Then the leading terms of coefficients in (\ref{g11})-(\ref{l22}) may be expressed as
\begin{eqnarray}
g_{11}&=&\left( 1+\kappa u\right) ^{2},\label{gbak11}\\
g_{12}&=&\left( z_{s}-u\tau \right) z_{u},\label{gbak12}\\
g_{22}&=&1+z_{u}^{2},\label{gbak22}\\
L_{11}&=&\frac{-z_{u}u\kappa ^{2}}{\sqrt{\left( z_{u}^{2}+1\right) }},\label{Lbak11}\\
L_{12}&=&\frac{z_{u}^{2}\tau }{\sqrt{\left( z_{u}^{2}+1\right) }}-\frac{z_{u}z\tau
}{\left( \frac{1}{\kappa }+u\right) \sqrt{\left( z_{u}^{2}+1\right) }},\label{Lbak12}\\
L_{22}&=&\frac{-z_{uu}}{\sqrt{\left( z_{u}^{2}+1\right) }}.\label{Lbak22}
\end{eqnarray}
We further obtain
\begin{eqnarray}
L_{12}g_{12} &=&\frac{\tau \left( z_{s}-u\tau \right) z_{u}^{3}}{\sqrt{%
\left( z_{u}^{2}+1\right) }}-\frac{z\tau \left( z_{s}-u\tau \right) z_{u}^{2}%
}{\left( \frac{1}{\kappa }+u\right) \sqrt{\left( z_{u}^{2}+1\right) }} \nonumber\\
&\sim &u\tau ^{2}z_{u}^{2}, \\
L_{11}g_{22} &=&\frac{-z_{u}u\kappa ^{2}\left( 1+z_{u}^{2}\right) }{\sqrt{%
\left( z_{u}^{2}+1\right) }}\sim u\kappa ^{2}z_{u}^{2}, \\
L_{22}g_{11} &=&\frac{-z_{uu}\left( 1+\kappa u\right) ^{2}}{\sqrt{\left(
z_{u}^{2}+1\right) }}\sim \frac{z_{uu}\left( 1+\kappa u\right) ^{2}}{z_{u}},
\end{eqnarray}%
which implies that
\begin{eqnarray}
L_{12}g_{12}\ll L_{11}g_{22},\quad L_{12}g_{12}\ll L_{22}g_{11}.
\end{eqnarray}
When writing the second term, we have used the argument $z_{uu}\sim\kappa z_u^3$ in Sec.~\ref{sec4}.
Besides, from (\ref{gbak11})--(\ref{gbak22}), we readily derive
\begin{eqnarray}
g_{12}^{2}\ll g_{11}g_{22}.
\end{eqnarray}
Thus the mean curvature may be expressed as
\begin{eqnarray}
2H &=&\frac{L_{11}g_{22}-2L_{12}g_{12}+L_{22}g_{11}}{g_{11}g_{22}-g_{12}^{2}}
\nonumber\\
&\approx &\frac{L_{11}g_{22}+L_{22}g_{11}}{g_{11}g_{22}} =\frac{L_{11}}{g_{11}}+\frac{L_{22}}{g_{22}}\nonumber\\
&\approx &\frac{-z_{u}}{\left( \frac{1}{\kappa }+u\right) \sqrt{\left(
1+z_{u}^{2}\right) }}-\frac{z_{uu}}{\left( 1+z_{u}^{2}\right) ^{\frac{3}{2}}}.
\end{eqnarray}
In addition, considering that
\begin{eqnarray}
L_{11}L_{22} &=&\frac{z_{u}z_{uu}u\kappa ^{2}}{\left( z_{u}^{2}+1\right) }%
\sim \frac{u\kappa ^{2}z_{uu}}{z_{u}+1}, \\
L_{12}^{2} &=&\left[ \frac{z_{u}^{2}\tau }{\sqrt{\left( z_{u}^{2}+1\right) }}%
-\frac{z_{u}z\tau }{\left( \frac{1}{\kappa }+u\right) \sqrt{\left(
z_{u}^{2}+1\right) }}\right] ^{2} \nonumber\\
&\sim &\tau ^{2}z_{u}^{2},
\end{eqnarray}
which implies $L_{11}L_{22}\gg L_{12}^{2}$, the Gaussian curvature may be expressed as
\begin{eqnarray}
K &=&\frac{L_{11}L_{22}-L_{12}^{2}}{g_{11}g_{22}-g_{12}^{2}} \approx\frac{L_{11}L_{22}}{g_{11}g_{22}} \nonumber\\
&\approx &\frac{z_{u}z_{uu}}{\left( \frac{1}{\kappa }+u\right) \left(
1+z_{u}^{2}\right) ^{2}}.
\end{eqnarray}

\section{Laplace operator}\label{secappG}

Considering the components of the metric (\ref{g11})-(\ref{g22}), we may obtain the first derivative of the components of the metric $g$ with respect to $u$
\begin{eqnarray}
g_{11u} &=&2\kappa \left( 1+\kappa u\right) +2\left( z_{s}-u\tau \right)
\left( z_{su}-\tau \right)  \nonumber\\
&&+2z\tau ^{2}z_{u}, \\
g_{12u} &=&\left( z_{s}-u\tau \right) z_{uu}+\left( z_{su}-\tau \right) z_{u},
\\
g_{22u} &=&2z_{u}z_{uu}
\end{eqnarray}
and $s$
\begin{eqnarray}
g_{11s} &=&2\left( 1+\kappa u\right) u\kappa _{s}+2\left( z_{s}-u\tau
\right) \left( z_{ss}-u\tau _{s}\right)  \nonumber\\
&&+2zz_{s}\tau ^{2}+2z^{2}\tau \tau _{s}, \\
g_{12s} &=&z_{s}\tau +z\tau _{s}+\left( z_{ss}-u\tau _{s}\right)
z_{u}+\left( z_{s}-u\tau \right) z_{su}, \\
g_{22s} &=&2z_{u}z_{su},
\end{eqnarray}
respectively.
In addition, the first derivative of the metric $g$ with respect to $u$ and $s$ can be obtained:
\begin{eqnarray}
g_{u} &=&g_{11u}g_{22}+g_{11}g_{22u}-2g_{12u}, \\
g_{s} &=&g_{11s}g_{22}+g_{11}g_{22s}-2g_{12s}.
\end{eqnarray}
Taking the analysis of the magnitude of the variables in last section into account, we know that $g_{22u}$ is much larger than other derivatives and thus $g_u\gg g_s$.

For function $h(s,u)$, the Laplace term takes the form of
\begin{widetext}
\begin{eqnarray}\label{laplaceapprox}
\nabla ^{2}h &=&\frac{1}{\sqrt{g}}\frac{\partial }{\partial u}\left( \frac{%
g_{11}}{\sqrt{g}}\frac{\partial h}{\partial u}-\frac{g_{21}}{\sqrt{g}}\frac{%
\partial h}{\partial s}\right) +\frac{1}{\sqrt{g}}\frac{\partial }{\partial s%
}\left( \frac{g_{22}}{\sqrt{g}}\frac{\partial h}{\partial s}-\frac{g_{12}}{%
\sqrt{g}}\frac{\partial h}{\partial u}\right)  \nonumber\\
&=&\frac{1}{\sqrt{g}}\left( \frac{g_{11u}\sqrt{g}-\frac{g_{11}}{2\sqrt{g}}%
g_{u}}{g}\frac{\partial h}{\partial u}-\frac{g_{12u}\sqrt{g}-\frac{g_{12}}{2%
\sqrt{g}}g_{u}}{g}\frac{\partial h}{\partial s}+\frac{g_{11}}{\sqrt{g}}\frac{%
\partial ^{2}h}{\partial u^{2}}-\frac{g_{21}}{\sqrt{g}}\frac{\partial ^{2}h}{%
\partial s\partial u}\right)   \nonumber\\
&&+\frac{1}{\sqrt{g}}\left( -\frac{g_{12s}\sqrt{g}-\frac{g_{12}}{2\sqrt{g}}%
g_{s}}{g}\frac{\partial h}{\partial u}+\frac{g_{22s}\sqrt{g}-\frac{g_{22}}{2%
\sqrt{g}}g_{s}}{g}\frac{\partial h}{\partial s}+\frac{g_{22}}{\sqrt{g}}\frac{%
\partial ^{2}h}{\partial s^{2}}-\frac{g_{12}}{\sqrt{g}}\frac{\partial ^{2}h}{%
\partial s\partial u}\right)   \nonumber\\
&\approx &\frac{1}{\sqrt{g}}\left( \frac{g_{11u}\sqrt{g}-\frac{g_{11}}{2%
\sqrt{g}}g_{u}}{g}\frac{\partial h}{\partial u}+\frac{g_{11}}{\sqrt{g}}\frac{%
\partial ^{2}h}{\partial u^{2}}+\frac{g_{22}}{\sqrt{g}}\frac{\partial ^{2}h}{%
\partial s^{2}}\right)
\end{eqnarray}
\end{widetext}
when ${\partial h}/{\partial u}\gg{\partial h}/{\partial s}$, ${\partial^2 h}/{\partial u^2}\gg{\partial^2 h}/{\partial s^2}$ and ${\partial^2 h}/{\partial u^2}\gg{\partial^2 h}/{{\partial s}{\partial u}}$.

According to equation (\ref{eq-2H-general}) in Sec~\ref{sec4}, i.e.,
\begin{equation}
2H= -\frac{\sin \psi }{\rho }-\frac{\partial  \sin \psi
 }{\partial u}%
,\quad K=\frac{\sin\psi}{\rho}\frac{\partial\sin\psi}{\partial u}
\end{equation}
and the assumption of $\psi\equiv\arctan {z_u}$ and $\rho\equiv u+1/\kappa(s)$, we may derive that
\begin{eqnarray}
&&\cos \psi  =\frac{1}{\sqrt{1+z_{u}^{2}}}, \quad \sin \psi =\frac{z_{u}}{\sqrt{%
1+z_{u}^{2}}} \\
&&\frac{\partial \rho }{\partial s} =\frac{-\kappa _{s}}{\kappa ^{2}}, \quad \frac{%
\partial \rho }{\partial u}=1 \\
&&\frac{\partial \psi }{\partial s} =\frac{\partial \left( \arctan
z_{u}\right) }{\partial s}=\frac{z_{us}}{1+z_{u}^{2}}, \quad \frac{\partial \psi }{%
\partial u}=\frac{z_{uu}}{1+z_{u}^{2}} \\
&&\frac{\partial ^{2}\psi }{\partial s^{2}} =\frac{z_{uss}}{\left(
1+z_{u}^{2}\right) }-\frac{2z_{u}z_{us}^{2}}{\left( 1+z_{u}^{2}\right) ^{2}},\\
&&\frac{\partial ^{2}\psi }{\partial u^{2}} =\frac{z_{uuu}}{\left(
1+z_{u}^{2}\right) }-\frac{2z_{u}z_{uu}^{2}}{\left( 1+z_{u}^{2}\right) ^{2}},\\
&&\frac{\partial ^{2}\psi }{\partial s\partial u}=\frac{z_{uus}}{\left(
1+z_{u}^{2}\right) }-\frac{2z_{u}z_{uu}z_{us}}{\left( 1+z_{u}^{2}\right) ^{2}%
}.
\end{eqnarray}
Thus
\begin{eqnarray}
\frac{\partial \left( 2H\right) }{\partial u} &=&-\frac{\cos \psi }{\rho }%
\frac{\partial \psi }{\partial u}+\frac{\sin \psi }{\rho ^{2}} \nonumber\\
&&+\sin \psi \left( \frac{\partial \psi }{\partial u}\right) ^{2}-\cos \psi
\frac{\partial ^{2}\psi }{\partial u^{2}}, \\
\frac{\partial ^{2}\left( 2H\right) }{\partial u^{2}} &=&\frac{\sin \psi }{%
\rho }\left( \frac{\partial \psi }{\partial u}\right) ^{2}+\frac{2\cos \psi
}{\rho ^{2}}\frac{\partial \psi }{\partial u}-\frac{\cos \psi }{\rho }\frac{%
\partial ^{2}\psi }{\partial u^{2}}  \nonumber\\
&-&\frac{2\sin \psi }{\rho ^{3}}+\cos \psi \left( \frac{\partial \psi }{%
\partial u}\right) ^{3}  \nonumber\\
&+&3\sin \psi \frac{\partial \psi }{\partial u}\frac{\partial ^{2}\psi }{%
\partial u^{2}}-\cos \psi \frac{\partial ^{3}\psi }{\partial u^{3}}, \\
\frac{\partial ^{2}\left( 2H\right) }{\partial s^{2}} &=&\frac{\sin \psi }{%
\rho }\left( \frac{\partial \psi }{\partial s}\right) ^{2}+\frac{2\cos \psi
}{\rho ^{2}}\frac{\partial \rho }{\partial s}\frac{\partial \psi }{\partial s%
}-\frac{2\sin \psi }{\rho ^{3}}\left( \frac{\partial \rho }{\partial s}%
\right) ^{2}  \nonumber\\
&+&\cos \psi \left( \frac{\partial \psi }{\partial s}\right) ^{2}\frac{%
\partial \psi }{\partial u}+\sin \psi \frac{\partial ^{2}\psi }{\partial
s^{2}}\frac{\partial \psi }{\partial u}  \nonumber\\
&+&2\sin \psi \frac{\partial \psi }{\partial s}\frac{\partial ^{2}\psi }{%
\partial u\partial s}-\cos \psi \frac{\partial \frac{\partial ^{2}\psi }{%
\partial s\partial u}}{\partial s}.
\end{eqnarray}%

The leading term of ${\partial \left( 2H\right) }/{\partial u}$, ${\partial ^{2}\left( 2H\right) }/{\partial s^{2}}$ and ${\partial ^{2}\left( 2H\right) }/{\partial u^{2}}$ is on the order of $\kappa^2/(1+u\kappa)$, $\kappa z_{su}^2/{z_u^2}$ and $\kappa^3 z_u^2/(1+u\kappa)$, respectively. By comparison of the highest orders of the remaining terms in (\ref{laplaceapprox}), the Laplace term in the free energy should be
\begin{eqnarray}
\nabla ^{2}\left( 2H\right)  &=&\frac{1}{\sqrt{g}}\left[ \frac{g_{11u}\sqrt{g%
}-\frac{g_{11}}{2\sqrt{g}}g_{u}}{g}\frac{\partial \left( 2H\right) }{%
\partial u}+\frac{g_{11}}{\sqrt{g}}\frac{\partial ^{2}\left( 2H\right) }{%
\partial u^{2}}\right]  \nonumber\\
&=&\frac{1}{\sqrt{g}}\frac{\partial }{\partial u}\left[ \frac{g_{11}}{\sqrt{g%
}}\frac{\partial \left( 2H\right) }{\partial u}\right].
\end{eqnarray}

% \end{widetext}

\end{document}